# Quantifying Precision in Measurement of Individualized, Evidence-Based Learning Across Sessions in Applied Behavior Analysis Therapy

Mark Louie F. Ramos[1], Reetam Majumder[2], Ji Young Kim[3], Jonathan W. Ivy[3]

[1]Department of Health Policy and Administration, The Pennsylvania State University

[2]Department of Mathematical Sciences, University of Arkansas

[3]School of Behavioral Sciences and Education, The Pennsylvania State University – Harrisburg

**Abstract**

Current performance criteria used in Applied Behavior Analysis (ABA) are often based on whether a learner meets a predetermined benchmark (e.g., 8 out of 10 correct responses across several sessions). Although these criteria help practitioners make instructional decisions, they provide limited information about how likely it is that the learner has truly acquired the skill. In this paper, we introduce a Bayesian approach that combines practitioners' prior expectations with the learner's ongoing performance data to estimate the learner's future performance level. Rather than providing only a pass/fail decision based on the criteria, we add an element of precision around that criteria to make decisions more nuanced to the testing conditions used. We also developed an R Shiny application for Measurement of Individualized, Evidence-Based Learning (MIEBL) that allows practitioners to apply this method in practice, supporting more informed instructional decisions and providing insight into the strengths and limitations of commonly used mastery criteria.

## Introduction

Performance criterion, also referred to as mastery criterion, is defined as a predetermined set of measurable performance standards that a learner must meet under specified conditions before a skill is considered acquired or mastered (McDougale et al., 2019; Richling et al., 2023). It is used in Applied Behavior Analysis (ABA) instructional approaches, such as in Discrete Trial Training (DTT), to inform clinical decision making that a student has performed a given skill to a sufficient level, demonstrating the skill is "learned" or "mastered" and thus allowing practitioners to modify, advance, or discontinue certain targets and programs. Because performance criterion directly affects instructional progression and skill acquisition, establishing performance criterion that practitioners can reliably count on is of critical importance.

Previous work introduced the concept of latent true mastery (LTM) of a skill, which is defined as the probability that a student will demonstrate that skill on some future response opportunity (Ramos, 2025a). The practitioner's interest is in deciding what the LTM might be after observing data that meets the prespecified performance criteria. Current practice implicitly equates LTM as being equivalent to the performance criterion used. For example, if a student responds 8 out of 10 trials for a skill correctly, then their LTM is assumed to be 80%. However, it is possible that the student achieves 60% or 100% even after demonstrating "mastery" during acquisition. This current practice underutilizes available data, which can provide more information around the uncertainty of that single estimate.

Maximizing the use of available data for deciding performance criteria is critical. If the criterion overestimates LTM, then practitioners will move on to training the next skill prematurely, which will negatively affect maintenance and generalization results. If the criterion underestimates true mastery, then practitioners may spend more time training a skill than necessary, thereby taking away time from learning new skills. These issues are well-recognized, and it has been repeatedly pointed out that typical performance criteria used in current practice tend to have variable outcomes during maintenance (Chang et al., 2024; Fienup & Carr, 2021; Kim et al., 2023; Longino et al., 2022; Richling et al., 2019). In other words, meeting the performance criterion during acquisition does not guarantee that a learner will perform at the same level in the future. While it is impossible to exactly ascertain the level of performance on some future task, previous work has attempted to model the uncertainty around the LTM attached to the choice of performance criterion using a simple beta-binomial Bayesian statistical framework (Ramos, 2025a).

Ramos (2025a) used a learner's pattern of correct and incorrect responses to estimate how confident a practitioner should be that the skill is truly learned or "mastered," rather than assuming that meeting a criterion automatically means the skill has been learned at that time. It studied the effect of different criterion, such as 8/10 correct responses versus 4/5, or 9/10 versus 8/10 on the LTM, and evaluated how different mastery criteria affect the probability that the learner will perform the skill correctly when given the next opportunity to respond. However, this model only considered the simple case of a single testing session. It was mainly meant to illustrate how different parameters in performance criterion, particularly the number of trials in the session and the number of correct responses required to pass, affected how much practitioners ought to believe that their student reached some level of mastery. In this paper, we extend this work to a more practical context where the performance criterion in ABA is typically implemented, namely through ongoing or continuous assessment. That is, the student is observed continuously from one session to the next until some performance criterion is met, such as performing 80% of trials correctly across three sessions in a row. We identify issues in modeling uncertainty around the student's LTM given their completion of a performance criterion in this context and present a prototype for making performance decisions that directly consider how the student's observed responses relate to their LTM instead of conditioning on a rigid performance criterion. We then constructed a functional prototype Rshiny web application for Measurement of Individualized, Evidence-Based Learning (MIEBL) that practitioners can use to implement our model. We address the following specific questions.

1. How does the choice of different criterion parameters (e.g., 80% versus 90% success per session, 2 consecutive versus 3 consecutive sessions, 5 trials per session versus 10 trials per session, etc.) change estimated probability (i.e., belief) about the student's mastery level after completing the criterion?
2. How can practitioners use the MIEBL app to help them make decisions about instruction?
3. How does using the MIEBL app conceptually compare to present practice?

**Performance Uncertainty Paradigm**

Previous work has pointed out that present practice involving the use of performance criterion in ABA does not formally account for the uncertainty around students' true mastery level upon accomplishment of the criterion (Ramos, 2025a). In this section, we formalize this performance uncertainty paradigm to be generally applicable to any performance criterion setting. A diagram for this paradigm is provided as Figure 1. The diagram shows two time periods in ABA therapy – training and maintenance. In between training and maintenance is acquisition, defined as the moment the student accomplishes the set performance criterion and is judged to have acquired the skill of interest. It is well known and accepted that just because a student has accomplished the performance criterion does not mean that the student has "mastered" the skill, where we define mastery here as the reliable emission of the target response when the relevant antecedent conditions are present (Fienup & Carr, 2021). But at the same time, it is well supported by literature and practice that students who accomplish some performance criteria are more likely to be able to demonstrate the skill than students who do not (Longino et al., 2022). Thus, it is more reasonable to think about "mastery" as the probability/chance that a student demonstrates the skill immediately after being observed to accomplish the performance criterion. This probability is denoted as predicted immediate maintenance (PIM). PIM is the performance probability before anything can change after the performance criterion was met. From here, the skill often enters a maintenance phase, where the intention is to check how well the skill has been retained and intervene as necessary. During the maintenance phase, it is well documented that the student's performance varies (Fienup & Carr, 2021). We denote predicted future maintenance ($PFM_t$) as the probability that the student will demonstrate the skill at a future time point t after acquisition, based only on information available when the performance criterion was met. Figure 1 provides a schematic of this uncertainty paradigm.

*Figure 1 Diagram for Performance Uncertainty Paradigm*

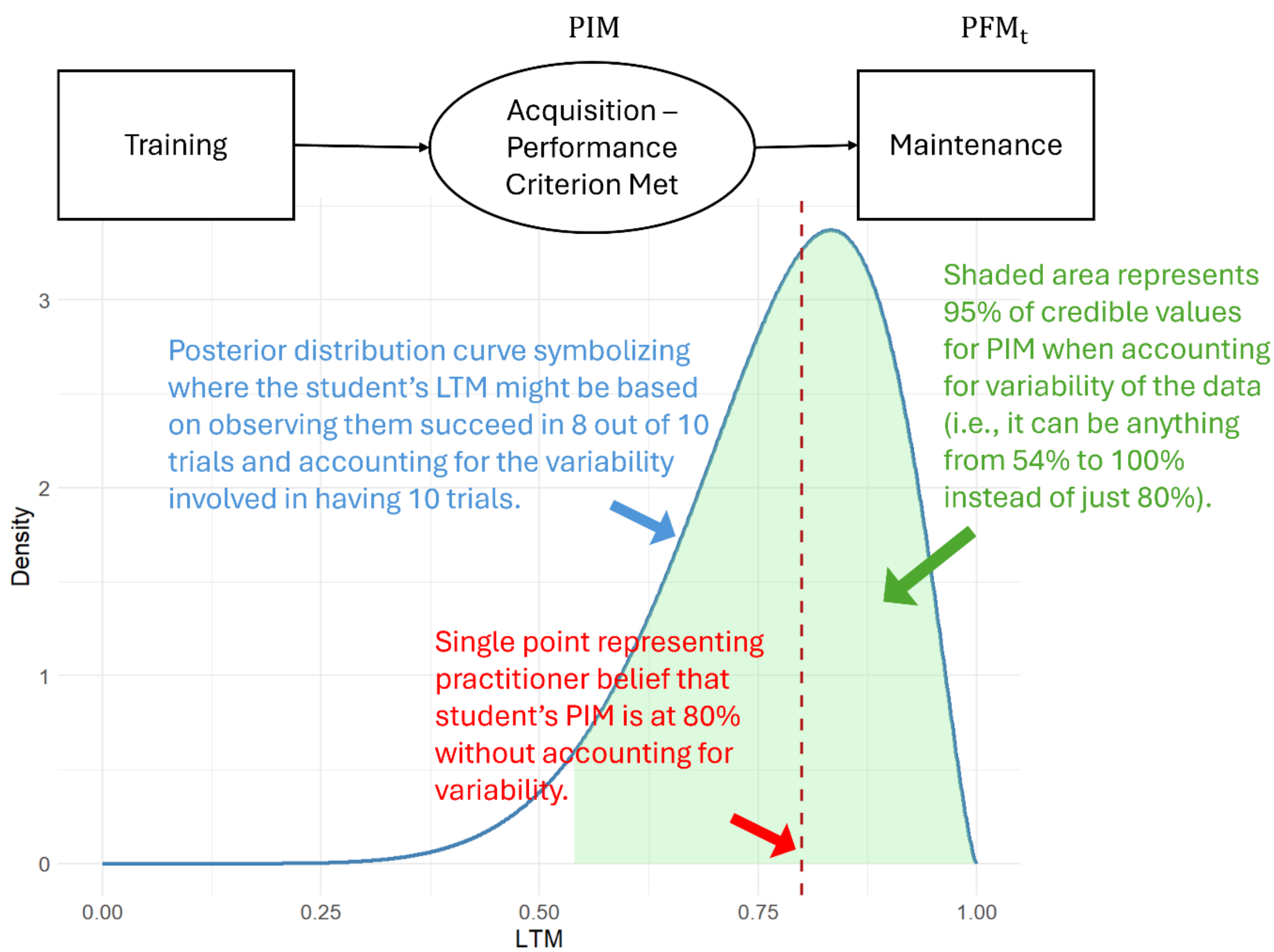

In this work, we extend the paradigm in Figure 1 to encompass the entire series of sessions around training a skill instead of just one session. This is visualized in Figure 2, which shows the familiar performance line graph used in practice and how our paradigm provides insight about where the student's LTM is both immediately after observing the desired performance criterion (PIM) and beyond ($PFM_t$). The hypothetical future performance can be drawn from the LTM with increasing uncertainty as we forecast further into the future. This can be expressed as a band as shown in Figure 2 or as a lower bound.

*Figure 2 Diagram for Multi-session Performance Uncertainty Paradigm*

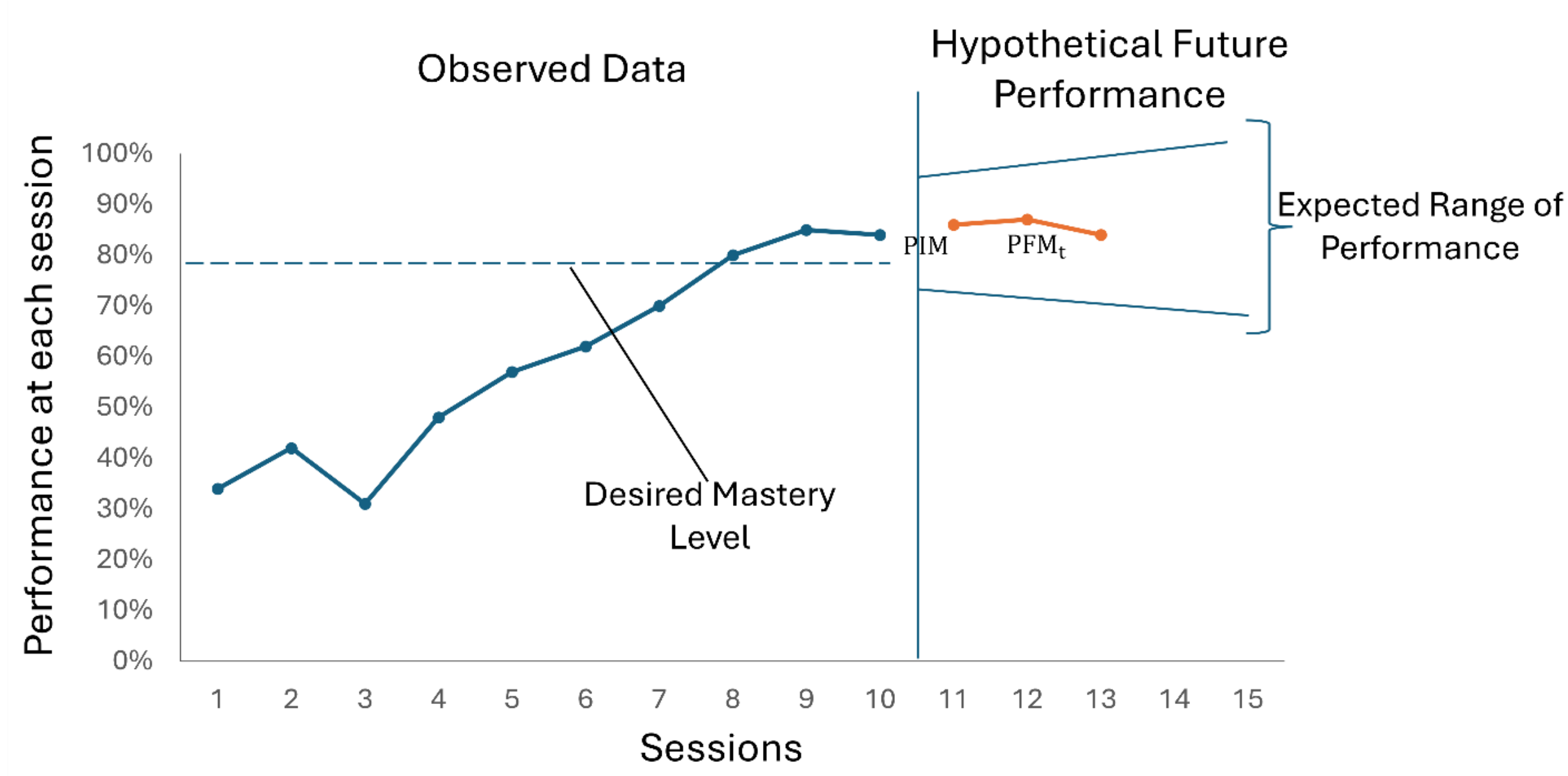

**Constructing reasonable posterior distributions for $LTM$**

Given the paradigm explained in the previous section, our objective is to construct what are called posterior distributions for LTM. A posterior distribution is a mathematical function that reflects our uncertainty about LTM suppose we observe some data, such as if the student accomplished the minimal requirement for a set performance criterion. This data is used to update our *a priori* understanding of a student's LTM, i.e., before any sessions are conducted. For example, the curve in Figure 1 is the posterior distribution for LTM after observing the student accomplish 8 out of 10 trials correctly. As more data is collected over several sessions, our *a posteriori* understanding of the LTM evolves. It is of technical importance to note here that the posterior distribution is constructed for LTM while PIM and $PFM_t$ are computed from that posterior distribution.

The mathematical construction of posterior distributions is an exercise in Bayesian statistics that we provide in detail in the supplementary materials, but it is sufficient to focus in this section on the elements of the performance criterion that shape the posterior distribution. First, we return to the single test setup where the only elements of the performance criterion are the number of trials in the session and the number of correct responses required to pass the criterion (Ramos, 2025a). Using software developed from this work (Ramos, 2025b), Figure 3 shows the posterior distributions for $LTM$ of students who respond 7, 8, or 9 out of 10 trials correctly in a single, independent session. This figure illustrates that a student who is observed to respond 7 out of 10 trials correctly may have $LTM$ that reasonably ranges from 30% to 90%, whereas the student who is observed to respond 9 out of 10 may have $LTM$ that is more reasonably within the ranges of 70% to 99%. This clearly shows the advantage of using 9 out of 10 instead of 7 out of 10 as the minimum performance criterion if the objective is to have confidence that the students who pass the criterion have a high enough level of mastery over the task.

*Figure 3 Posterior distributions of LTM for 10 trials*

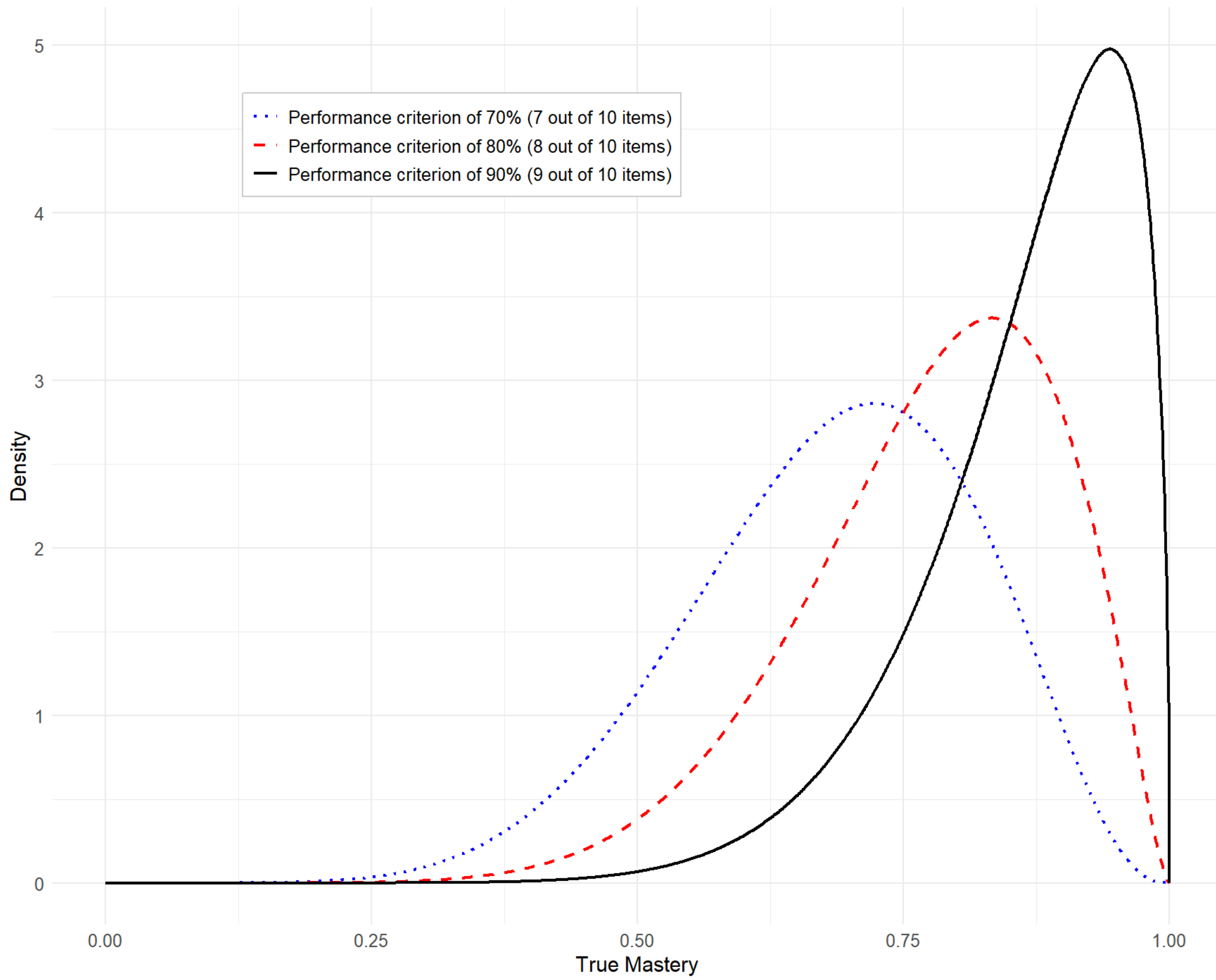

Similarly, Figure 4 shows the benefit of increasing the number of trials despite having the same performance criterion proportion, specifically for 80% performance criteria. Using 4 correct out of 5 total trials creates a posterior distribution with considerable variability, and reasonable values for $LTM$ that differ far from 80%, whereas using 16 correct out of 20 total trials creates a less variable posterior. This increased concentration of mass around the target level of mastery gives the practitioner greater confidence that the students mastery level does not differ far from 80%.

*Figure 4 Posterior distributions of LTM for 80% criterion*

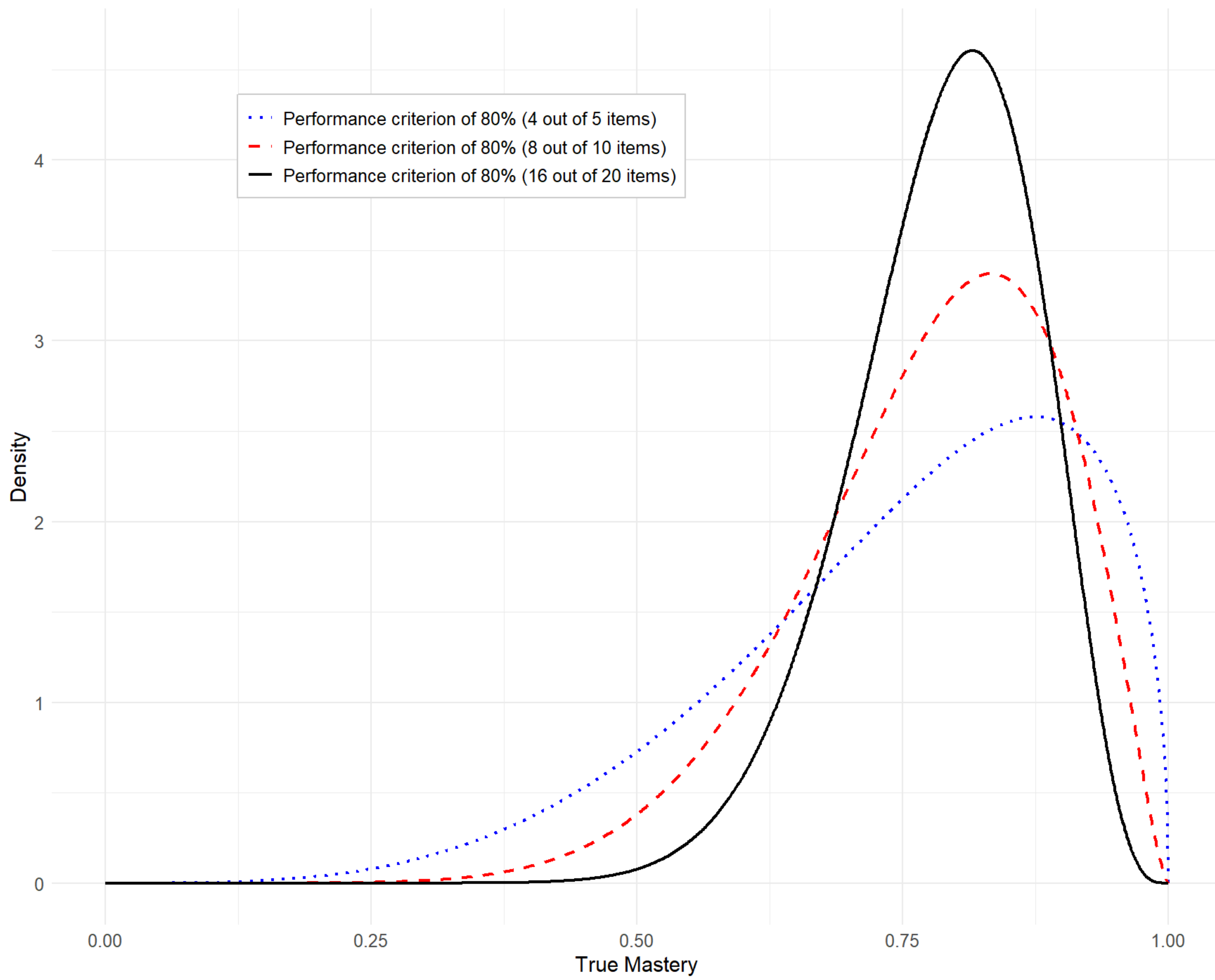

This simple approach is adequate when considering a single, pre-determined testing session following some learning period. It assumes that the trials can be reasonably viewed as identical and independent which is presently addressed by using inter-trial intervals (Booth & Keenan, 2018). However, in ABA practice, it is not typical to set a pre-determined testing session. Instead, the common practice is to continuously assess the student, session to session, until the student is observed to have performed at a prespecified performance level in some prespecified number of consecutive sessions (Fienup & Carr, 2021; Richling et al., 2023; Schneider et al., 2022), where common parameters used is 80% mastery in three consecutive sessions. This practice adds complexity that is not captured by Ramos (2025a). Since the student is assessed regularly across sessions, the model must account for the number of observed sessions prior to reaching the performance criterion. Two competing issues arise from this context.

First, we must incorporate the variability from observing the criterion at any latent mastery level as the number of sessions grows larger. That is, even if no learning was taking place and the student's true mastery level is far from adequate, there is a non-zero chance of observing accomplishment of the performance criterion and this chance increases as the window of observation becomes wider. For example, in matching activities where suppose a student has a 33% chance of selecting the correct response option by purely guessing, the chance that they will eventually do so correctly a consecutive number of times when given practically unlimited opportunities to do so is 100%. This means that only including the consecutive successful sessions in computing the posterior distribution for the latent mastery may result in overestimated upper-tail posterior probabilities. This is consistent with observations of maintenance performance variability among students for whom the same performance criterion was used (Fienup & Carr, 2021) and concerns about chance occurrence of criterion performance even when true mastery levels are low, which can come about due to unobserved issues like faulty stimulus control or irrelevant cues (Grow & LeBlanc, 2013).

Conversely, simply including the entire history of sessions for the student from the beginning of training poses the opposite practical challenge of potentially underestimating the latent mastery. Learning is a highly complex process and it is natural to expect that some amount of learning occurs from session to session. This means that the parameter estimated is the average latent mastery across the whole set of sessions observed rather than the specific latent mastery upon reaching the mastery level which is the desired parameter.

Ideally, for computing the posterior distribution, the practitioner should include as much of the latest data which best represent the student's present level of mastery. Ultimately, this is something that practitioners must decide on. In the following section, we present a model that is flexible to these decision criteria for constructing the posterior distribution of a student's latent mastery over a task as evidenced by a sequence of sessions.

**Multi-Session MIEBL app**

The app is available in Majumder et al., (2026) and Figure 5 shows its layout with an illustrative example using hypothetical data. The technical details in the computation of its outputs can be found in our supplementary materials. Key elements are labeled and described as follows.

A: Is the sequence of successful trials that the practitioner records per session. The app updates with each recorded session. In the example, the student has done 4 sessions so far and was successful in 6, 8, 9, and 8 trials respectively in each of those sessions.

B: Is the number of total trials attempted per session. If the practitioner intends to do a constant number of trials each session, as is typical, a single number can be entered instead of a sequence. In the example, each session has 10 trials. A sequence of 10's was still entered for illustration but a single 10 would have sufficed.

C: Is the per-session performance criterion expressed as percentage of successful trials per session. In the example, this is set to 80, which implies that a successful session is defined as one where the student is able to correctly accomplish at least 80% of the trials in that session.

D: Is the desired latent mastery level for the task $LTM'$. The practitioner sets this as their idea of what the acquisition level (not maintenance or generalization) of the student is sufficient for them to decide to move the student on from the task. For example, if this is set to 80, it means that if the practitioner wishes to be confident that the student can be expected to perform the task 80% of the time independently before they have the student move on.

E: Whereas D determines the desired latent mastery level, E is related to the practitioner's desired confidence level when computing PIM and $PFM_t$. The default is 0.95, which means that the intervals/bounds are constructed using 95% of the area under the posterior distribution.

F: This gives the lower bound for PIM from E. In the example, this means that the practitioner has 0.95 posterior probability belief that the student's acquisition level is such that they can do the task at least 70% of the time.

G: This computes the posterior probability for having at least the latent mastery indicated in D. That is, it computes $Pr(LTM \geq LTM')$. In the example, it shows that the posterior probability that the student's latent mastery is at least 80% is 0.66.

H: Is the prior distribution used for computations. This represents our initial belief of where the latent mastery is prior to observing the data. The present option, Jeffrey's prior, represents the fair situation where we either think that the student already knows the task sufficiently or completely does not know it. The influence of this choice gets weaker as more data is observed. More details about these options are in the supplementary materials.

I: Toggles whether to use the entire dataset entered or only the data points for sessions where the per-session criteria was achieved.

J: This shows the performance line graph with added PIM and $PFM_t$ values based on desired confidence level specified by the practitioner (default is 0.95 which encompasses 95% of values under the posterior distribution).

K: This shows the graph of the posterior distribution of LTM given the data entered.

L: Indicates how many future sessions is desired to be forecasted in J.

M: Indicates the type of forecasting method to be used. Technical details of each method is explained in the appendix.

The above are the core components needed to use the app. Practitioners can input their students' day-to-day data in the app and see how PIM and $PFM_t$ develops across sessions. They can also enter hypothetical data representing the present performance criterion they use and see how sensitive those are to the data in terms of values computed for F and G about the latent mastery.

*Figure 5: MIEBL App Interface*

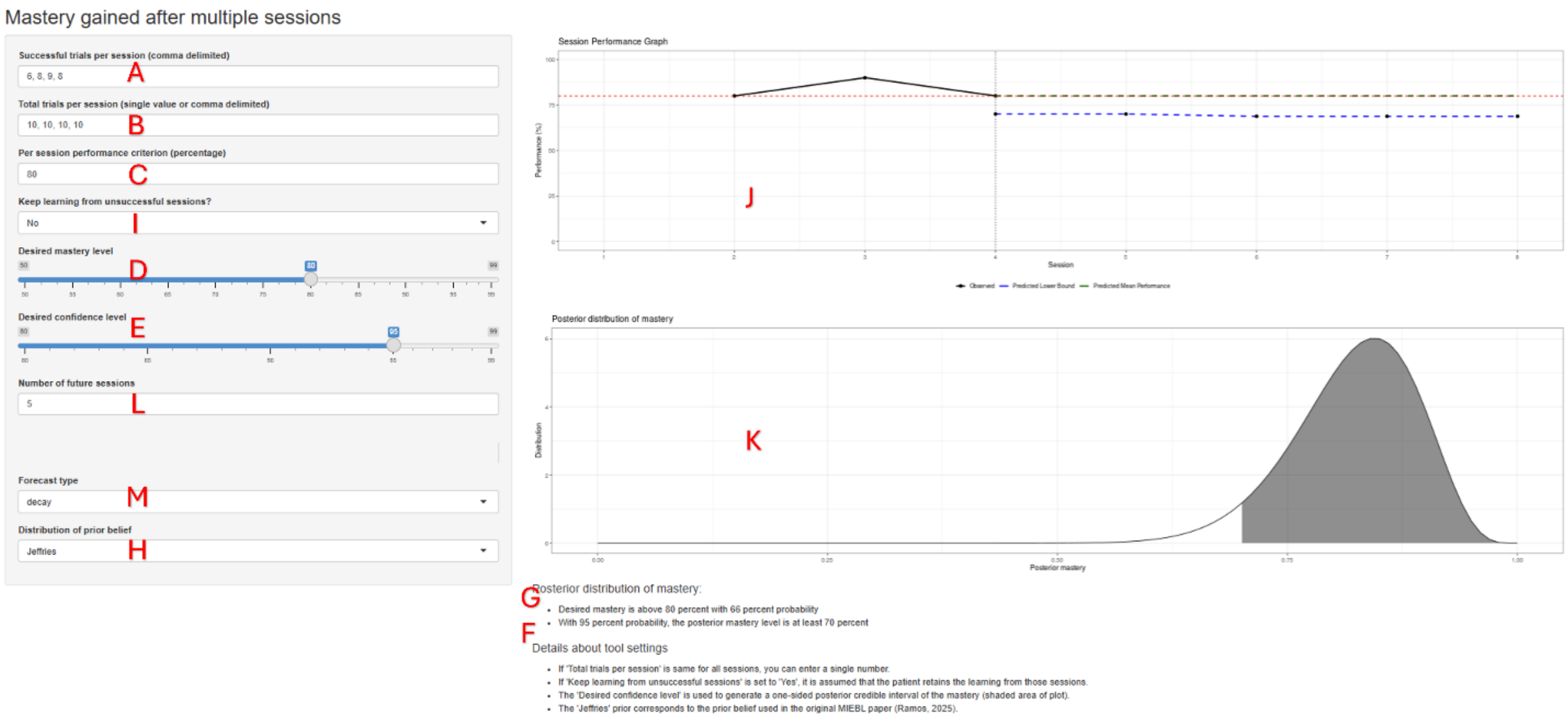

## Comparisons with Existing Performance Criteria

The primary advantage of including the app in a practitioner's repertoire of tools is that it allows the practitioner to probabilistically infer about the latent mastery level of the student, something that present performance criteria do not directly do. For example, when a practitioner sets a criterion of observing 80% performance in three consecutive sessions, there is an implicit claim made that meeting such a criterion means that the students' latent mastery is high enough to promote them to the next task. With the app, this claim is substantiated quantitatively using the data available. In Figure 5, the practitioner is made aware that information from the last 4 sessions implies that the student's latent mastery over the task is at least 70% with 0.95 probability, while the probability that it is above 80% is 0.66. The practitioner can then take this information and decide if they are good enough to promote the student, instead of just relying on the criteria. This provides greater insight for the practitioner, which can help substantiate decision-making where the criterion may feel too rigid. For example, consider if the observation in Figure 5 was 7, 10, 7, 10 instead. Under the 80% performance in three consecutive session criterion, that sequence would not pass, and yet the app computes that the student's latent mastery over the task is at least 74% with 0.95 probability, while the probability that it is above 80% is 0.78 – higher than the previous example. Objectively, based on the data alone, the student's performance in this example is better than in Figure 5, and it may make sense to the practitioner to pass the student in both situations.

The app also shows how using a static criterion over an unlimited horizon of observations creates issues we had previously described. For example, observing a sequence like 2, 2, 2, 2, 8, 8, 8 – the student was performing at 20% per session consistently and then suddenly does 3 consecutive sessions at 80%, can bring up potential issues of chance occurrence of criterion influenced by factors independent of acquisition (e.g., changes in instructors, time of the day, instructional materials used) (Fienup & Carr, 2021; Grow & LeBlanc, 2013). The app numerically substantiates such concerns. If we considered each session as reflective of the student's present mastery, then the app computes that the lower bound of the student's latent mastery with 0.95 probability is only 36% and there is 0 probability that the latent mastery is at least 80%. Conversely, the app requires the practitioner to be thoughtful about how much or which of the sessions to count when making a decision about the student's present level of mastery. Learning can occur from session to session, which means that the LTM and its distribution may change across sessions. Making the decision that the student has learned a skill is based on what we think the LTM is at the present as informed by data we have collected thus far. In current practice, when the practitioner decides this upon observing the student reach the criterion in three consecutive sessions, they are effectively only using those three sessions to make that decision. Our app allows the practitioner more flexibility by materially including more than these three sessions in the computation of the PIM and $PFM_t$ while still leaving the decision as to how far back of the data to include up to them. As such, practitioners can decide to use only the latest T sessions, discarding the oldest data point with data from each new session. Another option is to only include sessions that meet the per-session criterion in the computation of the posterior. This can be conveniently done in the app by toggling the option labeled under I in Figure 5 to "No." Another option is to set T as a fixed observation window, such that if the student does not have high enough values in F and G by the Tth session, the entire block is discarded for computation and observation begins anew. Practically, T in this sense represents the number of sessions elapsed before the practitioner decides that the present training method is not producing learning that will meet the criteria. While not usually set in terms of number of sessions, the use of rules for redirecting the instructional strategy is typically embedded in practice (Kipfmiller et al., 2019; Maffei-Almodovar et al., 2017).

## Conclusions and Ongoing Work

Making timely decisions on whether to promote the student to the next task or change instructional strategies is critical for effective learning in ABA therapy. Our proposed app can help practitioners do this by using Bayesian techniques to construct posterior probability distributions representing belief about the latent mastery level of the student over the task given observed data. The app is presently designed to be flexible to practitioner considerations about different elements of the observed sessions, including how many and which sessions to include for each instance of observation. More design options for the model, such as enabling iterative priors, are presently being considered. We hope to engage with practitioners directly to discuss how this tool might be customized to fit their practice.